\documentclass[aps,showpacs]{revtex4}          % onecolumn
\usepackage{graphicx}
\begin{document}

\title{On spacetime geometry above the electroweak symmetry breaking scale%\thanks{Grants or other notes
}


\author{Ka\'{c}a Bradonji\'{c}}
\affiliation{Department of Physics, Boston University, 590 Commonwealth Avenue, Boston, MA 02215}
\email[]{kacha@physics.bu.edu}
\date{March 24, 2011}

\begin{abstract}
While it is generally agreed that the nature of spacetime must be drastically different at the Planck scale, it has been a common practice to assume that spacetime is endowed with a full pseudo-Riemannian geometry regardless of the physical fields present or the length scale at which the geometry is probed, and to adopt this assumption in theories of cosmology, particle physics, quantum gravity, etc.  Following Einstein's view that the mathematical description of spacetime ought to be physically motivated, we initiate a discussion on the validity of this assumption, and propose that the full pseudo-Riemannian geometry of spacetime could emerge as late as the time of electroweak phase transition when spacetime acquires the projective structure necessary to describe the motions of massive particles.

\end{abstract}
\maketitle

\section{Introduction}
\label{intro}
Physics and mathematics have become inseparable in modern science. The clarity and the precision of mathematical language has been invaluable to understanding the nature of physical processes. But while it is expected that all physical laws have some mathematical formulation, it is generally accepted that not all mathematical models have a physical manifestation. Einstein recognized this asymmetric relationship and insisted that the geometrical description of spacetime be motivated by physical considerations.  Seeing the importance of the physical justification for the application of pure mathematical concepts to the description of natural phenomena, he emphasized a distinction between a ``purely axiomatic geometry" which rests on purely logical inferences and ``practical geometry" which is based on a principle accessible to experience \cite{Einstein1961}. In an essay based on a talk given at the Prussian Academy of Sciences in Berlin in 1922, Einstein clearly states that this distinction was crucial in the development of the general theory of relativity (GR). He further adds that ``\emph{[w]hether the structure of spacetime is Euclidian or Riemannian is a physical question which must be answered by experience, and not a question of a mere convention to be selected on practical grounds}" \cite{Einstein1922}. In fact, he goes as far as to say that the entire GR framework rests upon the assumption that two line segments defined on \emph{rigid bodies} equal at some time and place, are equal always and everywhere. 

The special and general theory of relativity, and consequently the attribution of pseudo-Riemannian geometry to spacetime, rest on physically meaningful notions of length, as that which is measured by a rigid rod,  and time, as that which which is measured by a physical clock.  Assuming that pseudo-Riemannian (or any other) geometry is an intrinsic property of spacetime, independent of the length scale or any changes in the matter fields present, is not justified.  In the same 1922 paper, Einstein himself questions the applicability of GR to sub-molecular scales and concedes that ``physical interpretation of geometry breaks down when applied immediately to spaces of sub-molecular order of magnitude." At the regime where the notion of ``rigid body" becomes meaningless, one must question the validity of any concept derived from such an object. 

The development of the Standard Model (SM) and the model of electroweak symmetry breaking (EWSB) opened up a possibility that there is an energy scale above which only massless particles exist \cite{Glashow,Weinberg,Salam,Donoghue}. Einstein did not live to witness these developments, and hence could not comment on any implications they would have on applicability of GR to currently interesting energy regimes. Most contemporary theories in cosmology, quantum gravity, and particle physics assume that the full pseudo-Riemannian geometry of spacetime still holds up to the Planck scale. Numerous approaches to the problem of quantum gravity start with a pre-spacetime structure in the hope of recovering the full pseudo-Riemannian manifold in some limit \cite{Thiemann,Rovelli,Fotini,Sorkin,Loll,Guth}. But if we consider Einstein's warning seriously, it becomes clear that the use of the full pseudo-Riemannian geometry as the accurate description of spacetime geometry when no massive particles are present, as it may be above EWSB scale according to the SM, is unwarranted unless some physical justification is provided.  

\section{Practical Geometry and the Standard Model}
\label{sec:1}

Today we know that relativity, along with its fundamental notions of length and time, is in fact applicable to sub-molecular regimes and to particle physics. This success is owed to the fact that the notions of length and time, although initially defined as measured by a rigid rod and a physical clock, can be retained at the particle level if they are redefined using massless and massive particles as measuring tools. As a matter of fact, Ehlers, Pirani, and Schild showed that a full pseudo-Riemannian spacetime geometry can be constructed from paths of massless and massive particles by imposing two physically motivated compatibility conditions between the two sets of paths \cite{Ehlers1972,Pirani1973}. One should, however, keep in mind Einstein's warning that, while such extrapolation may be possible, it should not be made without caution \cite{Einstein1922}:

\begin{quote}
``For even when it is a question of describing the electrical elementary particles constituting matter, the attempt may still be made to ascribe physical importance to those concepts of fields that have been physically defined for the purpose of describing the geometrical behavior of bodies that are large as compared with the molecule. Only the outcome can decide the justification of such an attempt, which postulates physical reality for the fundamental principles of Reimann's geometry outside of the domain of their physical definitions." 
\end{quote}

The enormous experimental success of the Standard Model, which assumes that spacetime is pseudo-Riemannian, provides the evidence Einstein would have demanded. The success is, however, restricted to energies for which experimental tests have been possible, and for which both massive and massless particles are present. While a pseudo-Riemannian manifold seems appropriate for modeling spacetime which accommodates the massless and massive particles, according to the SM, our universe underwent several phase transitions, one of which is the electroweak symmetry breaking \cite{Boyanovsky,Dawson}. According to the SM, prior to EWSB, the Higgs field had a vanishing vacuum expectation value (vev) and \emph{all the particles were massless}. If all physical particles are massless above EWSB scale, the set of available ``instruments" for measuring space and time is significantly reduced. We could then ask:
\begin{quote} ``Can we justifiably extrapolate the pseudo-Riemannain geometry to spacetime in the energy regimes where particles, \emph{as a matter of principle} and  \emph{in accordance with physical laws} do not have mass?"
\end{quote}
If we consider such regimes and take Einstein seriously, it is clear that the use of pseudo-Riemannain geometry as a description of spacetime, or standard GR,  in such circumstances ought to be questioned. This is a general question, and its relevance is \emph{independent} of the particular mechanism guiding the loss or generation of mass, or energy scale at which it happens. The Higgs mechanism is discussed here because it gives at least one simple and well-accepted model in which masses vanish, at least to a first approximation (neglecting interactions with thermal or quantum fluctuations). The electroweak scale is also interesting since it is possible to probe it experimentally.  One may object that particles, including Higgs boson, are never really massless due to radiative corrections which induce thermal mass terms and that, due to this induced mass, particles propagate along non-null geodesics in thermal plasma.  However, the framework of thermal field theory assumes that the full pseudo-Riemannian structure of spacetime is there in the first place. It also requires that one is working at finite temperature, which we need not do. Accessing a short distance scale, in a high energy collision for example, does not necessarily imply that one is working in a thermal background corresponding to that scale. In addition, one can always consider what happens at length scales which are below the mean free path of the collisions with the thermal background. The details of the Higgs mechanism do not eliminate the possibility that spacetime geometry may be different above the electroweak symmetry breaking scale, but below the Planck scale (where the manifold structure of spacetime itself is generally assumed to break down). This can only be verified or refuted empirically.

\section{Two Phases of Spacetime}
\label{sec:2}

We may ask what kind of properties we could attribute to spacetime if we had only massless fields at our disposal.  A hint as to what a geometry of spacetime at such a scale may be can be found in the previously mentioned work by Ehlers, Pirani, and Schild who have shown that a full pseudo-Riemannian geometry of spacetime can be ``built"  from structures defined by paths of massless and massive particles and by imposing two compatibility conditions between them \cite{Ehlers1972,Pirani1973}. The construction assumes a differentiable manifold and employs light rays (or free massless particles) and free massive particles as test bodies.  Here ``free"  means not under influence of anything but gravitational effects.  All the considerations of this formalism are local and assume that the manifold and the curves in question are differentiable. No consideration of the particle spin or other quantum numbers is made. In broad brushstrokes, the key steps of this axiomatic construction are:
\begin{itemize}
\item {The propagation of light determines at each point of spactime an infinitesimal null cone and hence defines a \emph{conformal structure} $\mathcal{C}$ on $M$. This structure allows for distinction among time-like, null, and space-like vectors, directions, curves, etc., at each point of $M$.  A conformal structure $\mathcal{C}$ on a manifold is mathematically described by an equivalence class  $\{\Omega^{2} g_{\mu\nu}\}$  of metrics $g_{\mu\nu}$ related by a positive factor $\Omega^{2}$, defined at every point of the manifold. Alternatively, $\mathcal{C}$ can be described by a tensor density field 
\begin{equation}
\tilde{g}_{\mu\nu}=(-g)^{-\frac{1}{4}}g_{\mu\nu},
\end{equation}
where $g$ is the determinant of the metric. $\mathcal{C}$ is invariant under the conformal transformations of the metric
\begin{equation}
g_{\mu \nu}\longrightarrow \Omega^{2} g_{\mu\nu},
\end{equation}
where $\Omega^{2}$ is a positive factor. Light rays are represented by null geodesics which are null curves contained in null hypersurfaces. }
\item{The motions of freely falling massive particles determine a family of preferred unparametrized time-like curves at each point, and such a family at each point of $M$ defines a \emph{projective structure} $\mathcal{P}$ on $M$. World lines of freely falling particles are said to be $\mathcal{C}$-time-like geodesics of $\mathcal{P}$. 
Projective structure is mathematically represented by projective parameters
\begin{equation}
\Pi^{\sigma}_{\rho \lambda}=\Gamma^{\sigma}_{\rho \lambda}-\frac{1}{5}(\delta^{\sigma}_{\rho}\Gamma_{ \lambda}+\delta^{\sigma}_{\lambda}\Gamma_{ \rho}),
\end{equation}
where $\Gamma_{\rho}=\Gamma^{\sigma}_{\rho \sigma}$ is the trace of the affine connection. $\mathcal{P}$ is invariant under the projective transformations of the affine connection
\begin{equation}
\Gamma^{\sigma}_{\rho \lambda} \longrightarrow \Gamma^{\sigma}_{\rho \lambda}+\frac{1}{5}(\delta^{\sigma}_{\rho}p_{\lambda}+\delta^{\sigma}_{\lambda}p_{\rho}),
\end{equation}
where $p_{\rho}$ is an arbitrary one-form. }
\item{Requiring two compatibility conditions: 1) that the null geodesics are also geodesics of $\mathcal{P}$, and 2) that the ticking rate of a clock is independent of its history, leads to the full pseudo-Riemannian geometry.}
\end{itemize}
The axiomatic construction of pseudo-Riemannian geometry from $\mathcal{C}$ and $\mathcal{P}$ and the two compatibility conditions illuminates the correspondence between the geometrical structures of spacetime and the physical fields, the behavior of which manifests these structures.  

Penrose already suggested that, as we go back close to the Big Bang, masses of all known particles could be considered ``effectively zero'',  and, assuming that particle interactions are conformally invariant, the early geometry is reduced to the conformal geometry \cite{Penrose2006}. According to this scenario, the particles never go to exactly zero mass and, at all stages, the full conformal and projective structures can be meaningfully defined. The argument presented here is different as we consider a phase where there are no massive particles at all and propose that such conformal geometry could happen at much lower energy.

If we restrict ourselves to an energy regime where there are no massive particles, the physically motivated axiomatic construction of \emph{practical} pseudo-Riemannian geometry of Ehlers, Pirani and Schild is not possible, as the absence of massive particles does not allow for the ``construction" of the projective structure. But in Einstein's own words,``[a]ccording to the general theory of relativity, the geometrical properties of space are not independent, but they are determined by matter" \cite{Einstein1961}. Since the matter fields in such a ``massless" regime cannot manifest the projective structure in any way, maintaining that $\mathcal{P}$, and hence the pseudo-Riemannian geometry, remains a property of spacetime is unjustified. 

Removing $\mathcal{P}$ from the pseudo-Riemannian spacetime leaves us with spacetime endowed with $\mathcal{C}$. Although such a spacetime is a mathematical possibility, it may not be physically meaningful. Even though ratios of lengths and angles, and conformal curvature are well defined in a conformally invariant spacetime \cite{Pirani1966}, it is an open question how, if at all, these structures are measurable, even in principle. 

These considerations lead to a conclusion that spacetime may have two geometrical phases: the first, possibly conformal phase, and the second, the familiar full pseudo-Riemannian phase. If this is correct, then, in the framework of the SM, spacetime at energies above EWSB is in the first phase. As electroweak symmetry phase transition gives particles their masses, massive particles give rise to the projective structure, hence breaking the conformal symmetry of spacetime and resulting in the emergence of a full pseudo-Riemannian geometry. This description could apply to the universe as a whole, or to a small region of spacetime with an energy content high enough to recreate the conditions prior to EWSB.  As previously noted, the considerations of this paper do not depend on the details of the process by which particles acquire or lose mass, but only assume that there is an energy above which particles are massless, and that they acquire mass below this energy. Whatever this energy may be is the energy at which we should question our assumption that spacetime is endowed with the full pseudo-Riemannian geometry.

\section{Conclusion}
\label{sec:3}
Maintaining the standards set by Einstein, we must admit that use of any geometry as a description of spacetime must be verified experimentally, or at least critically examined and physically justified. However, such examination seems to be lacking in most mainstream cosmology, quantum gravity, and particle physics models. It is the author's hope that this paper will initiate a more thorough and substantial discussion on the subject. If the proposal made here is true, there could be genuine new gravitational physics at the \emph{electroweak scale} possibly accessible to particle physics or cosmic ray experiments and cosmological observations right now \cite{Auger,LHC}.

\begin{acknowledgements}
I am grateful to the anonymous referee's thoughtful and constructive criticism of an earlier version of this paper which led me to add a note on the possible objections in the context of the Higgs mechanism. I would like to thank John Stachel for introducing me to the work on conformal and projective structures. I am also grateful to John Swain for reading an early draft of this paper, as well as for the discussion on possible physical consequences of scale dependent spacetime geometry. 
\end{acknowledgements}



\end{document}